\documentclass[11pt,fleqn]{article}
\usepackage{amssymb}

\newcommand{\half}{\mbox{$\frac{1}{2}$}}
\newcommand{\fslash}{\!\!\not\!}
\newcommand{\dslash}{\!\!\not\!\partial}
\newcommand{\tr}{\mbox{tr}}

\begin{document}

\begin{table}
\begin{flushright}
IK--TUW 9811401
\end{flushright}
\end{table}

\title{Jacobians of chiral transformations and \\
two-dimensional bosonization\thanks{Supported by Fonds zur
F\"{o}rderung der wissenschaftlichen Forschung, P11387--PHY}}

\author{Jan B. Thomassen\thanks{E-mail: {\tt
thomasse@kph.tuwien.ac.at}} \\
{\em Institut f\"{u}r Kernphysik, Technische Universit\"{a}t Wien} \\
{\em A--1040 Vienna, Austria}}

\date{February 3, 1999}

\maketitle

\begin{abstract}
We formulate a complete path integral bosonization procedure for any
fermionic theory in two dimensions. The method works equally well for
massive and massless fermions, and is a generalization of an approach
suggested earlier by Andrianov. The classical action of the bosons in
the bosonized theory is identified with $-i$ times the logarithm of
the Jacobian of a local chiral transformation, with the boson fields
as transformation parameters. Three examples, the Schwinger model, the
massive Thirring model and massive non-Abelian bosonization, are
worked out.

\vspace{\baselineskip}
\end{abstract}

\section{Introduction}

The derivation of the chiral Lagrangian of the strong interactions
\cite{gasser} from QCD is still an open question. This Lagrangian
describes the pseudoscalar octet -- the $\pi$'s, $K$'s and the $\eta$
-- in a way that is consistent with the scenario of spontaneously
broken chiral symmetry with the pseudoscalars as the Goldstone bosons.
Whatever the details of such a derivation may be, it is clear that one
of its ingredients must be bosonization, simply because the $\pi$'s,
$K$'s and $\eta$ are bosons.

There exists a physical and intuitively clear picture of how the
chiral Lagrangian arises from QCD [2--4]. In ref.\ \cite{diakonov}
Diakonov and Eides gave a formula for the classical action $W$ of a
given configuration of pseudoscalar mesons $\pi^a(x)$ in the chiral
limit. In our notation\footnote{Compared to refs.\
\cite{diakonov,balog}, the ratio in eq.\ (\ref{DE}) turns up side down
-- see below.},
\begin{eqnarray}
\label{DE}
e^{iW[\pi]} & = & \frac{\int{\cal D}B{\cal D}q{\cal D}\bar q
\exp i\left(S[B]+\int d^4x\bar q\;i\fslash\! D\; q\right)}
{\int{\cal D}B{\cal D}q{\cal D}\bar q
\exp i\left(S[B]+\int d^4x\bar qe^{i\Pi\gamma_5}\;i\fslash\! D\;
e^{i\Pi\gamma_5}q\right)},
\end{eqnarray}
where $B_\mu$ is the gluon field, $S$ is the gluon action, $q$ are the
$u$, $d$, and $s$ quarks and $\Pi\equiv\pi^at^a/f$, with $f=93 \mbox{
MeV}$ the pseudoscalar decay constant. It is understood that the
quarks both in the numerator and in the denominator are regularized
with a (Wess--Zumino consistent) vector current conserving
scheme. This renders the Jacobian $J[\pi]$ of a local chiral rotation
non-trivial, i.e.\ $W[\pi]$ is non-zero \cite{diakonov}, and we can
make the identification
\begin{eqnarray}
e^{iW[\pi]} & = & J[\pi].
\end{eqnarray}
The physical meaning of eq.\ (\ref{DE}) \cite{balog} is that the
classical action $W$ of the pseudoscalars is the difference in ``free
energy'' between chirally rotated QCD and QCD without this
rotation. The pseudoscalars, being the parameters of the chiral
rotations, are thus identified as Goldstone bosons of spontaneously
broken chiral symmetry.

Eq.\ (\ref{DE}) has been taken as the basis for deriving a chiral
Lagrangian from a cut-off version of QCD [2--4]. Furthermore, an
interesting connection between the underlying idea of this formula and
bosonization was pointed out \cite{andrianov,andrianov2}. In
particular, Andrianov noted that the Jacobian of a chiral rotation of
non-Abelian fermions realizes bosonization in two dimensions (in the
chiral limit) \cite{andrianov}. This motivates us to study this
particular approach to two dimensional bosonization.

Of course, the literature on two dimensional bosonization is
extensive, even if one restricts oneself to ``path integral
methods''. However, although these methods uses some of the same
ingredients (Jacobians of chiral transformations, identification of
the bosonic variables with parameters of chiral rotations, etc.) they
are essentially different when massive fermions are considered. For
example, some approaches uses path integrals in combination with
expansions in Green's functions. These can not be generalized to four
dimensions since Green's functions cannot in general be calculated
explicitly in that case. This mass-term problem justifies a systematic
treatment of Andrianov's scheme.

In this paper, then, we formulate a complete path integral
bosonization procedure for any fermionic theory in two dimensions,
based on Jacobians of chiral rotations. It is a generalization of the
bosonization procedure in the sense of Andrianov and can be used also
for fermionic models with mass-terms. That is, fermions coupled to
Abelian or non-Abelian $V$, $A$, $S$ and $P$ sources can be bosonized
with this procedure. The bosonization is complete in the sense that a
tensor is equivalent to a scalar in two dimensions.

The paper is organized as follows. In sec.\ 2 we formulate the
bosonization procedure and give a general bosonization formula. We
also discuss the sign of the action $W$. The ratio in eq.\ (\ref{DE})
is turned up side down compared to refs.\ \cite{diakonov,balog} in
order to produce the right sign \cite{eides,andrianov}, and we give a
physical argument for why the ratio should be turned the way it does
in eq.\ (\ref{DE}). We work out some examples in sec.\ 3 to
demonstrate the power and correctness of the procedure. In sec.\ 4 we
discuss some properties of the procedure and speculate on
generalizations to four dimensions. General expressions in Minkowski
space for the Jacobians of a chiral rotation are given in an appendix.

\section{The bosonization procedure}

Before we consider bosonization we give a physical interpretation of
the sign of $W$ in eq.\ (\ref{DE}). As we have mentioned, the sign
should be reversed \cite{eides,andrianov} compared to the expressions
in refs.\ \cite{diakonov,balog}.

A Grassmann integral over fermion fields describes a second quantized
system of fermions with a filled Dirac sea\footnote{See e.g.\ ref.\
\cite{jackiw} for a related discussion of the Dirac sea in the context
of chiral anomalies.}. The Dirac sea constitutes a {\em perturbative
vacuum} for the fermions but is not necessarily the ground
state. Thus, when the system is chirally rotated -- which disturbs the
Dirac sea -- it is perfectly possible that the ``free energy'' is {\em
lowered}, rather than increased. This is apparently what happens, at
least for QCD and the two-dimensional models we shall
consider. Therefore, if we wish to describe the pseudoscalars with a
{\em positive} ``free energy'' we must turn the ratio the way it does
in eq.\ (\ref{DE}).

Let us now turn to two dimensional bosonization. The basic formula
which we must consider is
\begin{eqnarray}
e^{iW[\theta]} & = & \frac{\int{\cal D}\psi{\cal D}\bar\psi
\exp i\int d^2x\bar\psi[i\dslash-\Gamma]\psi}
{\int{\cal D}\psi{\cal D}\bar\psi
\exp i\int d^2x\bar\psi e^{i\theta\gamma_5}[i\dslash-\Gamma]
e^{i\theta\gamma_5}\psi}
\end{eqnarray}
where $\psi$ is a Dirac fermion, $\Gamma=\;\;\fslash V+\;\fslash\!
A\gamma_5 +S+i\gamma_5P$ are external $V$, $A$, $S$ and $P$ sources
and $\theta$ is the pseudoscalar field -- the parameter of chiral
rotations. In the non-Abelian case $\psi$ is in the fundamental
representation of, say, $SU(N)$ and all quantities are matrices,
$\theta\equiv\theta^at^a$, etc.

The functional $W[\theta]$ is, as we have noted, the {\em classical}
action for the pseudoscalars. Thus, in order to get a {\em quantum}
theory, we must path integrate over $e^{iW[\theta]}$, leading to the
partition function of the bosonized theory:
\begin{eqnarray}
Z[V,A,S,P] & = & \int{\cal D}\theta e^{iW[\theta;V,A,S,P]}
\end{eqnarray}
where we have explicitly displayed the $V$, $A$, $S$ and $P$
dependence of $W$. The integration measure should be properly
normalized and have the appropriate invariance properties. For
example, in the non-Abelian case we should use the Haar measure. The
quantum theory of $\theta$ is not automatically regularized. That part
is left to the physicist who wishes to do calculations with the
bosonic theory.

The bosonization procedure is then as follows. Calculate the Jacobian
$J[\theta]=e^{iW[\theta]}$ of the finite local chiral rotation with
$\theta(x)$ as the rotation angles. Then the partition function of the
bosonized theory is found by path integrating $J[\theta]$ over
$\theta$, using the appropriate measure for the integration.

It is important that we use a consistent regularization scheme that
conserves vector currents in order to produce the correct action for
the bosons. More precisely, we should use the {\em most general}
regularization scheme with this requirement. We should not, for
example, choose specific simplifying values for the free parameters of
the scheme because we would then lose information. We will see
examples of this in the next section.

For convenience and completeness we give the full Minkowski space
expressions for the chiral Jacobians in an appendix.

\section{Examples}

We can now demonstrate that this bosonization procedure reproduces
known results in the cases of the Schwinger model, the massive
Thirring model and massive non-Abelian bosonization.

\bigskip
\noindent
{\em i) The Schwinger model} \cite{schwinger}

Due to the structural simplicity of this model, this example could be
regarded as a warm-up. The partition function is
\begin{eqnarray}
Z & = & \int{\cal D}A{\cal D}\psi{\cal D}\bar\psi
\exp i\int d^2x\left(-\mbox{$\frac{1}{4}$}F_{\mu\nu}^2
+\bar\psi[i\dslash-e\fslash\! A]\psi\right)
\end{eqnarray}
We choose Lorentz gauge for the photon field, $\partial_\mu
A^\mu=0$. Performing a chiral transformation of the fermion gives the
Jacobian
\begin{eqnarray}
J[\theta] & = & \exp i\int d^2x
\left(\frac{1}{2\pi}\partial_\mu\theta\partial^\mu\theta
-\frac{e}{\pi}A_\mu\epsilon^{\mu\nu}\partial_\nu\theta\right)
\end{eqnarray}
We bosonize the fermionic part of the partition function by replacing
it by this Jacobian, leading to
\begin{eqnarray}
\nonumber
Z & = & \int{\cal D}A\int{\cal D}\theta\exp i\int d^2x
\left(-\mbox{$\frac{1}{4}$}F_{\mu\nu}^2
+\frac{1}{2\pi}\partial_\mu\theta\partial^\mu\theta
-\frac{e}{\pi}A_\mu\epsilon^{\mu\nu}\partial_\nu\theta\right) \\
  & = & \int{\cal D}A\exp i\int d^2x
\left(-\mbox{$\frac{1}{4}$}F_{\mu\nu}^2+\frac{e^2}{2\pi}A^2\right),
\end{eqnarray}
which is appropriate for the description of a vector field with mass
$m=e/\sqrt{\pi}$.

\bigskip
\noindent
{\em ii) The massive Thirring model} [9--12]
%\cite{coleman}--\cite{damgaard}

We will only consider the mass-term bosonization rules for simplicity
and because it is the hardest part in the path integral
formalism. Current bosonization rules can be obtained in the same way
by coupling the Thirring fermion to external vector and axial vector
sources.

The partition function with external scalar sources $m(x)$ and
$m^\dagger(x)$
is
\begin{eqnarray}
\nonumber
Z[m,m^\dagger] & = & \int{\cal D}\psi{\cal D}\bar\psi\exp i\int d^2x
\left(\bar\psi[i\dslash-mP_+-m^\dagger P_-]\psi-\half gj^2\right) \\
  & = & \int{\cal D}B{\cal D}\psi{\cal D}\bar\psi\exp i\int d^2x
\left(\bar\psi[i\dslash-\:\fslash\! B-mP_+-m^\dagger P_-]\psi
+\frac{1}{2g}B^2\right)
\end{eqnarray}
where $P_\pm=\half(1\pm\gamma_5)$ and
$j_\mu=\bar\psi\gamma_\mu\psi$. The bosonization of the fermion is
given by the chiral Jacobian:
\begin{eqnarray}
\nonumber
J[\theta] & = & \exp i\int d^2x
\bigg(\frac{1}{2\pi}\partial_\mu\theta\partial^\mu\theta
-\frac{1}{\pi}B_\mu\epsilon^{\mu\nu}\partial_\nu\theta \\
  & & \mbox{} \hspace{6em} +\frac{1}{4\pi}\kappa_1m(e^{2i\theta}-1)
+\frac{1}{4\pi}\kappa_1m^\dagger(e^{-2i\theta}-1) \\
\nonumber
  & & \mbox{} \hspace{6em} +\frac{1}{8\pi}m^2(e^{4i\theta}-1)
+\frac{1}{8\pi}m^{\dagger 2}(e^{-4i\theta}-1)\bigg).
\end{eqnarray}
The coefficient $\kappa_1$ is an arbitrary mass which appears, for
example, in the Pauli--Villars scheme \cite{damgaard,ball}.

Bosonization is now completed by taking the path integral over
$\theta$:
\begin{eqnarray}
\nonumber
Z[m,m^\dagger] & = & \int{\cal D}B{\cal D}\theta\exp i\int d^2x
\bigg(\frac{1}{2\pi}\partial_\mu\theta\partial^\mu\theta
-\frac{1}{\pi}B_\mu\epsilon^{\mu\nu}\partial_\nu\theta
+\frac{1}{2g}B^2 \\
\nonumber
  & & \mbox{} \hspace{10em} +\frac{1}{4\pi}\kappa_1m(e^{2i\theta}-1)
+\frac{1}{4\pi}\kappa_1m^\dagger(e^{-2i\theta}-1) \\
  & & \mbox{} \hspace{10em} +\frac{1}{8\pi}m^2(e^{4i\theta}-1)
+\frac{1}{8\pi}m^{\dagger 2}(e^{-4i\theta}-1)\bigg) \\
\nonumber
  & = & \int{\cal D}\varphi\exp i\int d^2x
\bigg(\half\partial_\mu\varphi\partial^\mu\varphi
+\frac{1}{4\pi}\kappa_1m(e^{i\beta\varphi}-1)
+\frac{1}{4\pi}\kappa_1m^\dagger(e^{-i\beta\varphi}-1) \\
\nonumber
  & & \mbox{} \hspace{9em} +\frac{1}{8\pi}m^2(e^{2i\beta\varphi}-1)
+\frac{1}{8\pi}m^{\dagger 2}(e^{-2i\beta\varphi}-1)\bigg)
\end{eqnarray}
In the second expression we have rescaled $\theta$,
\begin{eqnarray}
\varphi & = & \sqrt{\frac{1}{\pi}\left(1+\frac{g}{\pi}\right)}\;\theta,
\end{eqnarray}
and introduced the parameter $\beta$ by
\begin{eqnarray}
\label{coupling}
\frac{4\pi}{\beta^2} & = & 1+\frac{g}{\pi}.
\end{eqnarray}

We are allowed to add suitable polynomial counterterms in $m$ and
$m^\dagger$ to the action. If we choose
\begin{eqnarray}
{\cal L}_{ct} & = & \frac{1}{4\pi}\kappa_1(m+m^\dagger)
+\frac{1}{8\pi}(m^2+m^{\dagger 2})
\end{eqnarray}
we can read off the bosonization rules:
\begin{eqnarray}
\label{rules}
\nonumber
-\sigma_+ & = & \frac{\kappa_1}{4\pi}e^{i\beta\varphi}
+\frac{1}{4\pi}me^{2i\beta\varphi}, \\
-\sigma_- & = & \frac{\kappa_1}{4\pi}e^{-i\beta\varphi}
+\frac{1}{4\pi}m^\dagger e^{-2i\beta\varphi},
\end{eqnarray}
with $\sigma_\pm\equiv\bar\psi P_\pm\psi$. The presence of the $m$-
and $m^\dagger$-dependent terms makes this a ``field dependent''
bosonization rule \cite{divecchia}. Modulo these terms -- which
generate contact terms in the Greens functions -- eqs. (\ref{rules})
are the usual mass bosonization rules \cite{coleman,dorn}.

We emphasize that we should not choose any special values for
$\kappa_1$. In particular, we should not choose $\kappa_1=0$, since we
would then lose the sine in the sine--Gordon equation of the bosonic
theory. A similar arbitrary mass is also present in Coleman's
expressions \cite{coleman}.

\bigskip
\noindent
{\em iii) The massive non-Abelian case} \cite{dorn,polyakov,witten}

Here too we will only consider mass-term bosonization -- bosonization
of the currents, as we have already mentioned, is covered by the
remark in ref.\ \cite{andrianov}. The partition function is
\begin{eqnarray}
Z[m,m^\dagger] & = & \int{\cal D}\psi{\cal D}\bar\psi
\exp i\int d^2x\left(\bar\psi[i\dslash-mP_+-m^\dagger P_-]\psi\right),
\end{eqnarray}
where $\psi$ is now an $SU(N)$ multiplet and $m=m^at^a$,
$m^\dagger=m^{\dagger a}t^a$ are in the Lie algebra. From the chiral
rotation we get the Jacobian
\begin{eqnarray}
\nonumber
J[U] & = & \exp i\int d^2x\bigg(\frac{1}{8\pi}
\tr\:\partial_\mu U^\dagger\partial^\mu U
+\frac{1}{12\pi}\int_0^1dt\:\epsilon_{\mu\nu\tau}
\tr\left(\hat U^\dagger\partial^\mu\hat U
\hat U^\dagger\partial^\nu\hat U
\hat U^\dagger\partial^\tau\hat U\right) \\
  & & \mbox{} \hspace{6em} +\frac{1}{4\pi}\kappa_1\tr\:m(U-1)
+\frac{1}{4\pi}\kappa_1\tr\:m^\dagger(U^\dagger-1) \\
\nonumber
  & & \mbox{} \hspace{6em} +\frac{1}{8\pi}\tr\,[mUmU-m^2]
+\frac{1}{8\pi}\tr\,[m^\dagger U^\dagger m^\dagger
U^\dagger-m^{\dagger 2}]\bigg)
\end{eqnarray}
where $U=e^{2i\theta}$, $\hat U=e^{2it\theta}$, and our
three-dimensional conventions are $x^2\equiv t$, $g_{\mu\nu}=(+,-,-)$
and $\epsilon_{012}=1$. Again, this leads to the required form for the
bosonized theory \cite{witten} modulo contact terms.

\section{Discussion and speculations}

These examples demonstrates clearly that in two dimensions the
procedure of calculating the Jacobian of a chiral rotation in the most
general vector current conserving regularization scheme and then path
integrating over it realizes bosonization. This bosonization procedure
was suggested earlier by Andrianov \cite{andrianov} in the special
case of massless non-Abelian fermions. In this paper we have displayed
the full generality of the procedure.

Of course, only the information in the ``uncharged sector'' of the
fermionic theory can be cast in a bosonized form, corresponding to
Greens functions obtainable from the partition function by
differentiation wrt.\ $V$, $A$, $S$ and $P$ sources. ``Fermionic
information'' -- that which is expressable in terms of Feynman
diagrams with external fermion lines -- can not be obtained from a
bosonic theory.

Let us also point out that in a non-Abelian model we could also
bosonize the singlet, or $U(1)$, degree of freedom in addition to the
$SU(N)$ ones. The complete Jacobian is then just a product of the
singlet and non-singlet Jacobians, and mixed Greens functions in the
bosonized theory can be found accordingly.

One may ask why the bosonization procedure works since we are unable
to actually {\em derive} the bosonic form of the partition function
from its fermionic form. Under a chiral rotation the partition
function transforms such that
\begin{eqnarray}
Z[\Gamma] & = & e^{iW[\theta;\Gamma]}Z[\theta;\Gamma]
\end{eqnarray}
where $Z[\theta;\Gamma]$ is the rotated partition function without the
Jacobian. Since, due to bosonization,
\begin{eqnarray}
Z[\Gamma] & = & \int{\cal D}\theta\;e^{iW[\theta;\Gamma]}
  \;=\; \int{\cal D}\theta\frac{Z[\Gamma]}{Z[\theta;\Gamma]}
\end{eqnarray}
we must have \cite{andrianov}
\begin{eqnarray}
\int{\cal D}\theta\frac{1}{Z[\theta;\Gamma]} & = & 1.
\end{eqnarray}
It is necessary to prove this equation if we want to ``prove'' the
bosonization procedure, but no such proof is known to us. The
procedure {\em does} work, however, since it reproduces the
bosonization rules that have been proved by other methods, for example
by explicitly comparing the Greens functions in the fermionic and
bosonic theories, respectively.

We can speculate that the chiral degrees of freedom somehow saturate
the theory in two dimensions. This will also imply that the same
procedure can only be approximately correct in four. If there are
other symmetries of a fermionic theory in four dimensions that
produces a Jacobian when the partition function is locally
transformed, then the degrees of freedom associated with this
transformation would also be important for bosonization. In fact,
attempts at ``bosonizing'' QCD [2--4] according to eq.\ (\ref{DE})
using {\em only} chiral degrees of freedom gives a chiral Lagrangian
with a quadratically divergent ``kinetic term'' for the pions. This
violates the principle that physics should not depend on the
regularization scheme. The inclusion of further degrees of freedom
could be part of a cure for this problem.  More generally, it appears
to us not altogether improbable that if we could identify all such
symmetries of a fermionic theory, we would be able to find a
completely bosonized version of this theory.

\noindent
\paragraph{Acknowledgments} I would like to thank M. Faber,
P.H. Damgaard and A.N. Ivanov for discussions and comments on the
manuscript, and M.I. Eides, also for comments on the manuscript.

\section*{Appendix}

In this appendix we give complete expressions for the Jacobians of
chiral rotations in Minkowski space. The partition function is
\begin{eqnarray}
Z & = & \int{\cal D}\psi{\cal D}\bar\psi\exp i\int d^2x
\bar\psi[i\dslash\;-\fslash V-\;\fslash\! A\gamma_5-S-i\gamma_5P]\psi
\end{eqnarray}
where all fields are either Abelian or non-Abelian. We are interested
in the Jacobian of the change of variables
\begin{eqnarray}
\psi & = & e^{i\theta\gamma_5}\chi, \hspace{2em}
\bar\psi \;=\; \bar\chi e^{i\theta\gamma_5}.
\end{eqnarray}
In the Abelian case, the Jacobian is \cite{dorn,damgaard}
\begin{eqnarray}
\nonumber
J[\theta] & = & \exp i\int d^2x
\bigg(\frac{1}{2\pi}\partial_\mu\theta\partial^\mu\theta
-\frac{1}{\pi}V_\mu\epsilon^{\mu\nu}\partial_\nu\theta
+\frac{1}{\pi}A_\mu\partial^\mu\theta \\
  & & \hspace{6em} \mbox{} +\frac{1}{4\pi}\kappa_1m(e^{2i\theta}-1)
+\frac{1}{4\pi}\kappa_1m^\dagger(e^{-2i\theta}-1) \\
\nonumber
  & & \hspace{6em} \mbox{} +\frac{1}{8\pi}m^2(e^{4i\theta}-1)
+\frac{1}{8\pi}m^{\dagger 2}(e^{-4i\theta}-1)\bigg)
\end{eqnarray}
where $m\equiv S+iP$. In the non-Abelian case we get \cite{dorn,pak}
\begin{eqnarray}
\nonumber J[U] & = & \exp i\int d^2x
\bigg(\frac{1}{8\pi}\tr\:\partial_\mu U^\dagger\partial^\mu U
+\frac{1}{12\pi}\int_0^1dt\:\epsilon_{\mu\nu\tau}
\tr\left(\hat U^\dagger\partial^\mu\hat U
\hat U^\dagger\partial^\nu\hat U
\hat U^\dagger\partial^\tau\hat U\right) \\
\nonumber
  & & \hspace{4em} \mbox{} -\frac{1}{4\pi}\tr
\Big[R_\mu i\partial^\mu UU^\dagger
-U^\dagger i\partial_\mu UL^\mu
-\epsilon_{\mu\nu}(R^\mu i\partial^\nu UU^\dagger
-U^\dagger i\partial^\mu UL^\nu) \\
  & & \hspace{8em} \mbox{} +U^\dagger R_\mu UL^\mu-R_\mu L^\mu
-\epsilon_{\mu\nu}(U^\dagger R^\mu UL^\nu-R^\mu L^\nu)\Big] \\
\nonumber
  & & \hspace{4em} \mbox{} +\frac{1}{4\pi}\kappa_1\tr\: m(U-1)
+\frac{1}{4\pi}\kappa_1\tr\: m^\dagger(U^\dagger-1) \\
\nonumber
  & & \hspace{4em} \mbox{} +\frac{1}{8\pi}\tr\,[mUmU-m^2]
+\frac{1}{8\pi}\tr\,[m^\dagger U^\dagger m^\dagger U^\dagger
-m^{\dagger 2}]\bigg)
\end{eqnarray}
where we use standard $LR$-notation: $R_\mu=V_\mu+A_\mu$ and
$L_\mu=V_\mu-A_\mu$. For the coefficient $\kappa_1$ and the other
conventions, see the text.

\end{document}